\documentclass[a4paper,11pt,5p]{elsarticle}
\usepackage{siunitx}
\usepackage{multirow}		%Needed for multi-row cells in tables
\usepackage{ctable}			%Needed for \toprule etc in table
\usepackage{threeparttable}
\biboptions{sort&compress}	
\begin{document}

\begin{frontmatter}

	\title{The Interaction of Galling and Oxidation in 316L Stainless Steel}

		\author[IC]{Samuel R. Rogers\corref{cor1}}
		\ead{srr13@ic.ac.uk}
		\author[MAN,AEA]{David Bowden}
		\author[MAN]{Rahul Unnikrishnan}
		\author[MAN]{Fabio Scenini}
		\author[MAN]{Michael Preuss}
		\author[RR]{David Stewart}
		\author[IC]{Daniele Dini}
		\author[IC]{David Dye}
		\cortext[cor1]{Corresponding author}
		\address[IC]{Imperial College, South Kensington, London SW7 2AZ, UK}
		\address[MAN]{Department of Materials, The University of Manchester, Sackville Street Building, Manchester M1 3BB, UK}
		\address[AEA]{UK Atomic Energy Authority, Culham Science Centre, Abingdon OX14 3DB, UK}
		\address[RR]{Rolls-Royce plc, Raynesway, Derby DE21 7WA, UK}

	\begin{abstract}
		The galling behaviour of 316L stainless steel was investigated in both the unoxidised and oxidised states, after exposure in simulated PWR water for \SI{850}{\hour}. Galling testing was performed according to ASTM G196 in ambient conditions. 316L was found to gall by the wedge growth and flow mechanism in both conditions. This resulted in folds ahead of the prow and adhesive junction, forming a heavily sheared multilayered prow. The galling trough was seen to have failed through successive shear failure during wedge flow. Immediately beneath the surface a highly sheared nanocrystalline layer was seen, termed the tribologically affected zone (TAZ). It was observed that strain-induced martensite formed within the TAZ. Galling damage was quantified using R$_{t}$ (maximum height - maximum depth) and galling area (the proportion of the sample which is considered galled), and it was shown that both damage measures decreased significantly on the oxidised samples. At an applied normal stress of \SI{4.2}{\mega\pascal} the galled area was \SI{14}{\%} \textit{vs.} \SI{1.2}{\%} and the R$_{t}$ was \SI{780}{\micro\meter} \textit{vs.} \SI{26}{\micro\meter} for the unoxidised and oxidised sample respectively. This trend was present at higher applied normal stresses, although less prominent. This difference in galling behaviour is likely to be a result of a reduction in adhesion in the case of the oxidised surface.

	\end{abstract}

\end{frontmatter}

\section{Introduction}
	
	Cobalt-based hardfacing alloys are used in nuclear applications on account of their high wear and galling resistance. Under the ALARA (as low as reasonably achievable) principle \cite{ONRCobalt}, cobalt must be removed from nuclear applications. Cobalt is not used in the reactor pressure vessel (RPV) and so components do not undergo direct irradiation. However, after extended use, components wear, with wear debris travelling into the RPV, becoming irradiated and transmutating from \textsuperscript{59}Co to \textsuperscript{60}Co, which is a gamma radiation source. Since the wear debris will continue to travel around the primary circuit, it may cause additional doses to personnel working on and around the primary circuit, including during shutdowns.  As such, alternative Co-free materials are desired for tribologically sensitive components such as valve seats.
	
	Austenitic stainless steels containing hard particles have been suggested for some time as replacement materials for the Stellite\textsuperscript{TM} family of alloys (Co-Cr-W with W- and Cr-carbides), which are currently the most widely used cobalt alloys used in nuclear applications. In both cases, the alloy matrix is fcc austenite, with the ability to form strain-induced martensite (hcp $\epsilon$-martensite in Stellite\textsuperscript{TM} and bct $\alpha'$-martensite in austenitic stainless steels). A number of galling resistant stainless steel alloys have been developed over the past four decades which incorporate martensite formation during wear \cite{RTOcken,NOREM02Friction,BowdenThesis,GallTough,SteelersGalling,HardfacingIndustrialPractices}. However, none have been considered suitable for wide-scale use in reactors, owing to their reduced galling resistance at elevated temperature, such as those seen in light water reactors. Further work is therefore necessary to develop a stainless steel which is galling resistant at elevated temperatures.
	
	Galling is an adhesive wear mechanism, active at slow sliding speeds and relatively high compressive stresses. It is reported to cause gross plastic deformation of mated surfaces, particularly when their movement is bound \cite{TribologyBook}.
	
	A number of works have investigated the mechanisms of galling and their relation to surface deformation. Some concluded that adhesion and galling occurs primarily through the agglomeration of wear particles and that these heavily work-hardened particles adhere to one surface and gouge the opposing surface \cite{Cocks1958,Antler1962,Antler1964,Sasada1976}. Other works have concluded that galling appears to occur through the adhesion of opposing asperities which shear to failure, and may also result in the formation of peaks and troughs \cite{Cocks1962,Cocks1964}. Through this mechanism, layering of material has been observed to occur, resulting in the formation of the peaks and the formation of `lips' in subsequently formed troughs \cite{Glaeser1981,StickSlip,SuccessiveWear}.
	
	Although work has predominantly been focussed upon the mechanism of surface deformation and failure, some work has also been carried out on the sub-surface changes observed after adhesive wear and galling.  A number of authors have reported the formation of a heavily sheared sub-surface region \cite{SmithThesis,TribologyBook} which, for austenitic stainless steels has been found to contain strain-induced $\alpha'$-martensite (SIM) \cite{SmithThesis,KimKim}. As such, martensite is widely considered to be a source of galling resistance in stainless steels since the reduction in galling resistance correlates with the reduction in SIM formation at elevated temperatures \cite{KimKim}. This heavily sheared region is similar in appearance to the sub-surface microstructural changes observed after fretting, termed the white layer \cite{TTSFrettingMechanisms,TTSFrettingSteel}.
	
	Much of the work on galling in the literature has been concerned with the \emph{qualification} of galling, with little work being produced on the \emph{quantification} of galling. Examples of this include the ASTM G98 and G196 galling tests which state whether a sample has or has not galled at a given load, in order to find a threshold galling load, ASTM G98 \cite{ASTMG98}, or the proportion of samples which gall at a given load (galling frequency), ASTM G196 \cite{ASTMG196}. Budinski and Budinski sought to improve the recording of results for these tests by introducing a scoring system, corresponding to the type of damage seen \textit{e.g.} burnishing, adhesive transfer and incipient galling, however, these results are not strictly speaking quantitative \cite{GallingMatrix}.  Ives \textit{et al.} significantly developed the quantification of galling, using the average maximum peak-to-valley height, root-mean-square of R$_{t}$, displaced volume and damage aspect ratio to quantify a single galled sample \cite{GallingBook}.
	
	An area of research which has not been widely explored is the galling behaviour of an oxidised metal substrate, despite observations which suggest significant improvement of galling resistance in simulated light water reactor conditions \cite{KimKim}. This knowledge gap is addressed in this work.
	
	Many galling resistant stainless steels were developed from a base composition of 316 or 304 stainless steel, with the addition of large volume fractions of hard phases (carbides, nitrides or silicides). Here, 316L stainless steel is used to investigate the galling behaviour of a stainless steel matrix material in both the bare-metal and oxidised states, without the complication of ceramic hard phase additions.

\section{Method}
	
	%\subsection{}
			316L bar, supplied by Goodfellow, was manufactured into ASTM G196 specimens with R$_{t}$ = \SI{10}{\micro\meter} (maximum height - maximum depth), and machined with the surface lay circumferential.

	%\subsection{Oxide Growth}
		An autoclave was used to produce a representative PWR environment, enabling a representative oxide to be formed on galling specimens before testing. 316L stainless steels were oxidised at \SI{300}{\degreeCelsius} for \SI{850}{\hour} in a static autoclave, with a water pressure of \SI{120}{\bar}. The water chemistry was controlled to contain 2 ppm Li, which was added as LiOH,10.5 pH, 4 ppm dissolved $H_{2}$ and less than 5 ppb of $O_{2}$.

	%\subsection{Galling Test}
	
			\begin{figure}[h]
				\centering
				\includegraphics[width=9cm]{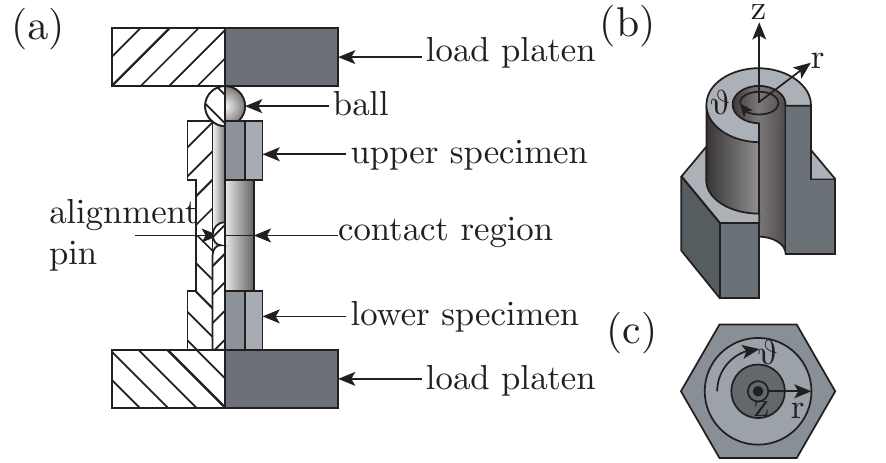}
				\caption{(a) ASTM G196 galling rig, redrawn from \cite{ASTMG196}; (b) ASTM G196 galling sample, with a section removed to an enable a view of the radial cross-section; (c) top view of an ASTM G196 galling sample.}
				\label{Rig}
			\end{figure}	
	
		An ASTM G196 rig, Figure \ref{Rig}, was used to perform the galling tests. All tests were self-mated, in either the oxidised or unoxidised state, and performed at ambient temperature and pressure. A torque of \SI{350}{\newton\meter} was applied using a torque wrench, taking approximately \SI{60}{\second} to complete a single revolution. The normal stress was applied using a hydraulic loading cylinder, controlled to \SI{4}{\mega\pascal} -- \SI{103}{\mega\pascal} (the lower limit of the equipment, and a representative contact stress for gate valves in nuclear power plant, respectively \cite{CoLWR}). If a test pair seized, the test was finished when seizure occurred. If seizing occurred, the adhesive junction was broken before the mating surfaces could be observed.
		
		Before testing, the mating surfaces were cleaned using propanol. After testing, surfaces were left undisturbed.
		
	%\subsection{Galling Quantification}
			A Veeco white light interferometer and an Olympus OLS5000 LEXT confocal microscope were used to detect surface topography and generate sample surface reconstructions.
		
		Post-processing was employed to remove surface artefacts, sample edges and to generate data which was missing due to a lack of light detection. Linear interpolation of nearest neighbours was used to remove sample artefacts and reconstruct the full sample surface. In addition, surfaces were translated such that the minimally worn and ungalled regions were considered flat and at the zero plane (\SI{0}{\micro\metre} in height).
				
		 A number of galling measures developed and used by Ives \textit{et al.} will be used in this work; the maximum height, depth and R$_{t}$. In addition to these, the galled area was calculated, where the galled area is the proportion of the sample which is either above or below a threshold height value, corresponding to the initial surface R$_{t}$.

	%\subsection{Microstructural Examination}
		Samples were prepared for metallographic examination by grinding through to 4000 grit SiC paper, using a diamond suspension as a first polishing stage, and a final polishing stage using an OPU suspension. Imaging was then produced using Zeiss Sigma 300 and Auriga SEM's in both secondary (SE) and backscattered electron (BSE) modes. XFlash 6160 and Oxford Instruments X-Max detectors in the Sigma and Auriga SEM's were also used for imaging and X-ray microanalysis.
		
		In addition, a FEI Helios Ga FIB/SEM was used to perform site-specific \textit{in-situ} lift outs for observation in a JEOL 2100Plus TEM. The TEM also contained STEM and STEM-EDX capabilities, which were used in conjunction with a Panalytical X'pert Pro diffraction (XRD) system in order to investigate oxide chemistries and structures and the fine sub-surface microstructural features seen as a result of galling.

		Phase identification was performed using the JEOL 2100 Plus TEM in diffraction mode and a Panalytical X'pert Pro diffraction (XRD) system.

\section{Results \& Discussion}
\subsection{Oxide Characterisation} 
	
				\begin{figure*}[h]
				\centering
				\includegraphics[width=18cm]{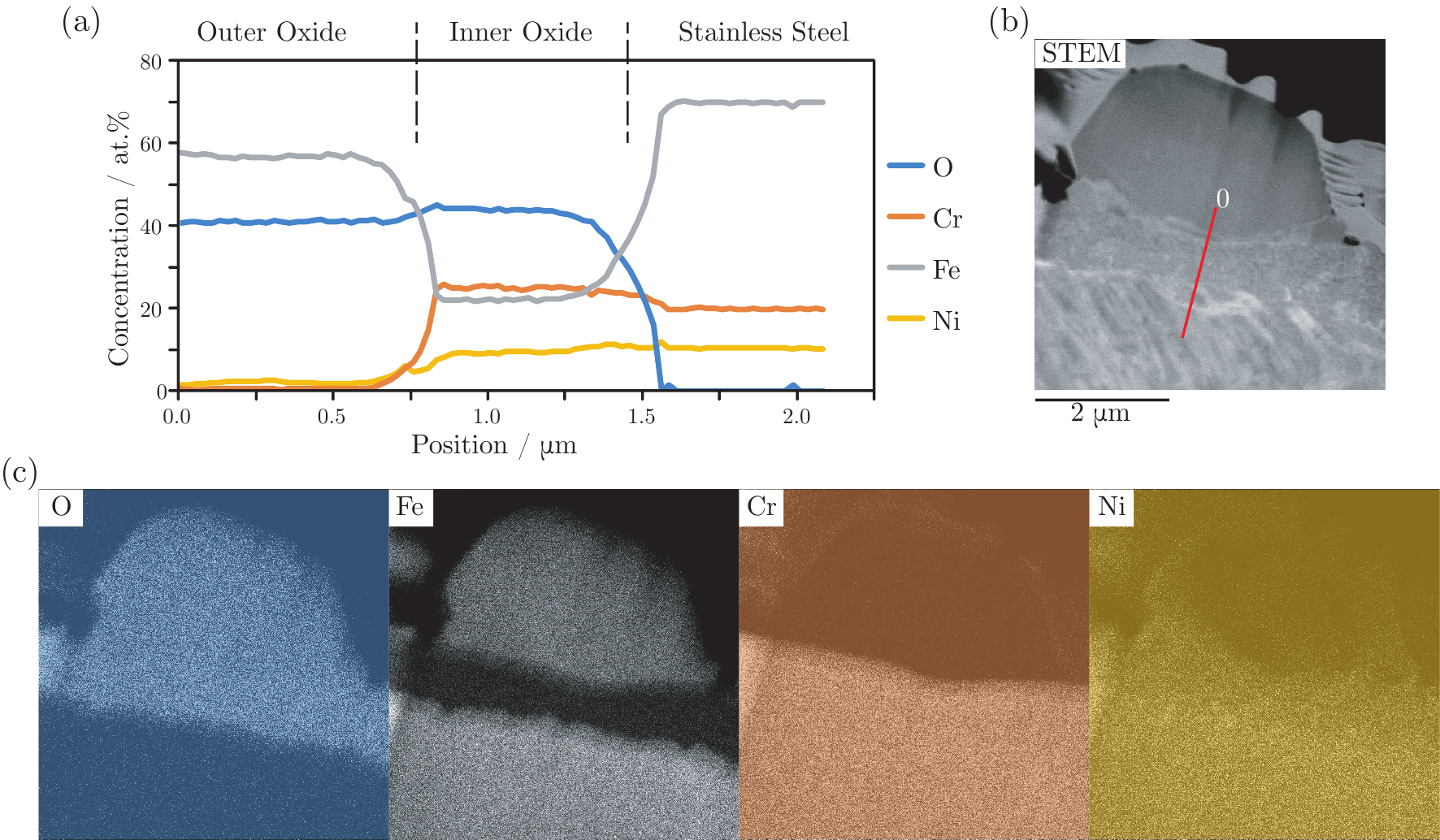}
				\caption{(a) The concentration profiles produced using STEM-EDX for the four primary elements in the oxides produced through autoclaving of 316L stainless steel. (b) A STEM image showing a cross-section of the oxides and the position of the STEM-EDX line scan. (c) EDX elemental maps for the region shown in (b).}
				\label{OxideSTEM-EDX}
			\end{figure*}
			
		An \textit{in-situ} lift-out was taken from an oxidised surface and shows that two oxide layers are formed on the surface of 316L after autoclaving in simulated PWR conditions for \SI{850}{\hour}, Figure \ref{OxideSTEM-EDX}. The outer oxide layer can be seen to be made up of discrete crystallites, whilst the inner oxide was nanocrystalline and free of pores and voids at the metal-oxide interface.
		
		Compositional information, found using STEM-EDX, showed that the outer oxide is Fe-rich and depleted of both Cr and Ni. In contrast, the inner oxide is Cr-rich as well as containing both Fe and Ni, Figure \ref{OxideSTEM-EDX}. 
		
		Structural information was found using XRD, showing that a single oxide structure was present,  M$_{3}$O$_{4}$ (where M is a metal ion), often known as magnetite. The oxide peaks appeared as doublets, suggesting that both the inner and outer oxides have the same structure, but with differing lattice parameters, likely as a result of their different chemistries. The outer oxide was therefore found to be Fe-rich magnetite of composition Fe$_{3}$O$_{4}$, whilst the inner oxide is a Cr-rich magnetite of approximate composition Cr$_{1.3}$Fe$_{1.2}$Ni$_{0.5}$O$_{4}$, in agreement with Terachi \textit{et al.} \cite{CorrosionPWRWaterChemistry} and Kim \cite{Kim316}.

	\subsection{Macroscopic Observations}
		
			\begin{figure}[h]
				\centering
				\includegraphics[width=9cm]{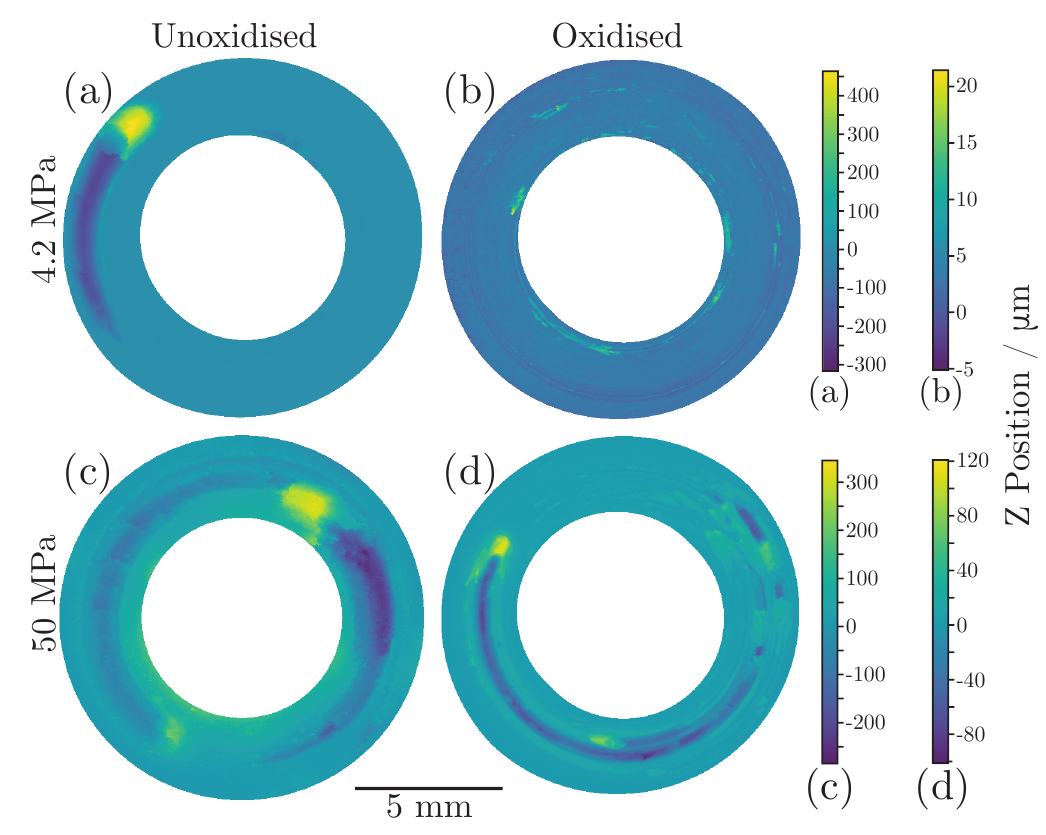}
				\begin{threeparttable}
				\centering
				%\caption{Quantification of the galling damage using surface profile measurements for samples shown in Figure \ref{WLI} .}
				\label{MeasuringGalling}
				%\begin{tabular}{m{3.5em}m{6em}m{2.5em}m{2.5em}m{2.5em}} %l=left justified; m=middlealigned; >{\raggedright} for not justifying text - USE THIS ONE TO HAVE MORE CONTROL OVER COLUMN SIZES (at least, it works...)
				{\small
				\begin{tabular}{ccccc}
				\toprule
				\toprule
					\multirow{2}{3.5em}{\textbf{Sample}}     & \multirow{2}{5em}{\textbf{Galled Area / \%}} & \multicolumn{3}{c}{\textbf{Sample Height / \SI{}{\micro\meter}}}	\\
					&	&    \textbf{Max}   & \textbf{Min}   & \textbf{Rt}         \\
					
				\midrule
					(a)   & 14                 & 460                &     -320     &       780       \\
					(b)   & 1                 & 21                &     -5     &       26       \\
					(c)   & 59                 & 350                &     -280     &       630       \\
					(d)   & 28                 & 120                &     -100     &       220       \\
				\bottomrule
				\end{tabular}
				}
			\end{threeparttable}

				\caption{White light interferometry height maps of 316L samples in the unoxidised (a) \& (c), and oxidised (b) \& (d) conditions, with their corresponding heights and height scales.}
				\label{WLI}
			\end{figure}

			\begin{figure}[h]
				\centering
				\includegraphics[width=9cm]{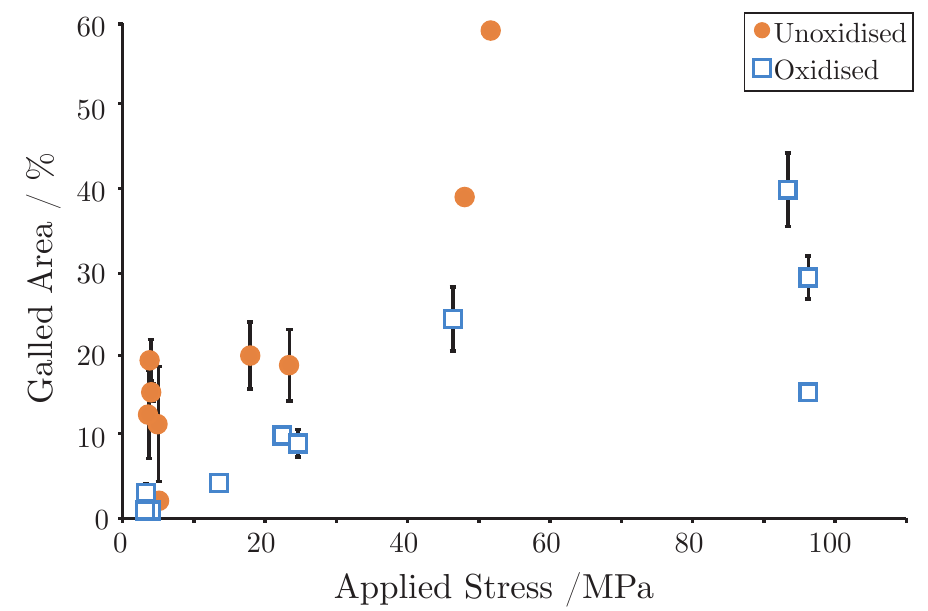}
				\caption{The effect of oxidation on galling for 316L stainless steel using galled area as a measure of galling severity. Error bars represent the difference between each surface of a single galled pair. Where these are small, the error bar is smaller than the marker.}
				\label{Quant}
			\end{figure}

		In the unoxidised condition, 316L stainless steel was seen to gall at the lowest applied stresses of \SI{3.8}{\mega\pascal}. As a result, it was felt that the ASTM G98 standard concept of a threshold galling stress was not a suitable measure of galling, since it is a purely qualitative measure and would not differentiate between the extent of galling damage seen across different samples.	 As a result, quantitative measures developed by Ives \textit{et al.} have been used in this work.
		
		In the unoxidised condition at an applied normal stress of \SI{4.2}{\mega\pascal} it was observed that 316L formed a single galling peak and trough with a sample R$_{t}$ of \SI{780}{\micro\meter} and a galling area of \SI{14}{\%}, Figure \ref{WLI}. It was seen that when the applied normal stress was increased to \SI{50}{\mega\pascal} the galled area was seen to increase, as expected, however, the sample R$_{t}$ decreased to \SI{640}{\micro\meter} when compared with the sample tested at \SI{4.2}{\mega\pascal}. Since two samples are required for the galling tests, by observing the damage on the other sample within the test pair, it was seen that the average R$_{t}$ for the test pair galled at \SI{50}{\mega\pascal} was in fact larger than that of the test pair galled at \SI{4.2}{\mega\pascal}. There was, however, large variability in the R$_{t}$ in unoxidised samples tested across the applied normal stress range making a conclusion regarding the effect of applied normal stress on R$_{t}$ difficult. 
		
		In contrast, the oxidised 316L stainless steel samples were seen to behave as expected, since both the R$_{t}$ and galled area were seen to increase with increased applied normal stress across the full range of applied normal stresses, Figures \ref{WLI} \& \ref{Quant}.
						
		For a given applied normal stress, the extent of galling seen by the unoxidised samples was considerably greater than that of the oxidised samples, Figure \ref{WLI}. For the samples tested at a low applied stress, the R$_{t}$ was also on a different order of magnitude when comparing 316L in the unoxidised and oxidised conditions; \SI{780}{\micro\meter} \textit{vs.} \SI{26}{\micro\meter}. Although a less pronounced difference was observed with regard to R$_{t}$, at high applied stresses, there was again a significant difference in the galled area, with the galled area being considerably larger when tested in the unoxidised condition; \SI{14}{\%} \textit{vs.} \SI{1.2}{\%}, Figure \ref{Quant}. This is unsurprising, since it is well known that adsorbed oxygen in ambient conditions gives rise to adhesion resistance when compared with adhesion under vacuum, and an oxide layer is a surface film which is more wear resistant than adsorbed oxygen  \cite{BuckleyHex,BuckleySummary}.
				
		The most crucial finding was that for one of the unoxidised tests at \SI{50}{\mega\pascal}, seizure occurred before the end of the test, with the opposing surfaces needing to be pulled apart for observation.
				
		It was also be noted that the number of galling prows on a surface appears to be larger on the oxidised samples than the unoxidised samples, where typically there are only one or two galling peaks, Figure \ref{WLI}. This suggests that either multiple galling instances take place simultaneously, or that prow growth is interrupted by oxide, and so abrasion recommences until metal-metal adhesion and subsequent galling can re-occur. The morphology of the galling scars was consistent throughout the tests, despite the change in damage magnitude. This behaviour suggests that the same mechanism was active for both the oxidised and unoxidised samples.

	\subsection{Galling Mechanisms}
		\begin{figure}[h]
			\centering
			\includegraphics[width=9cm]{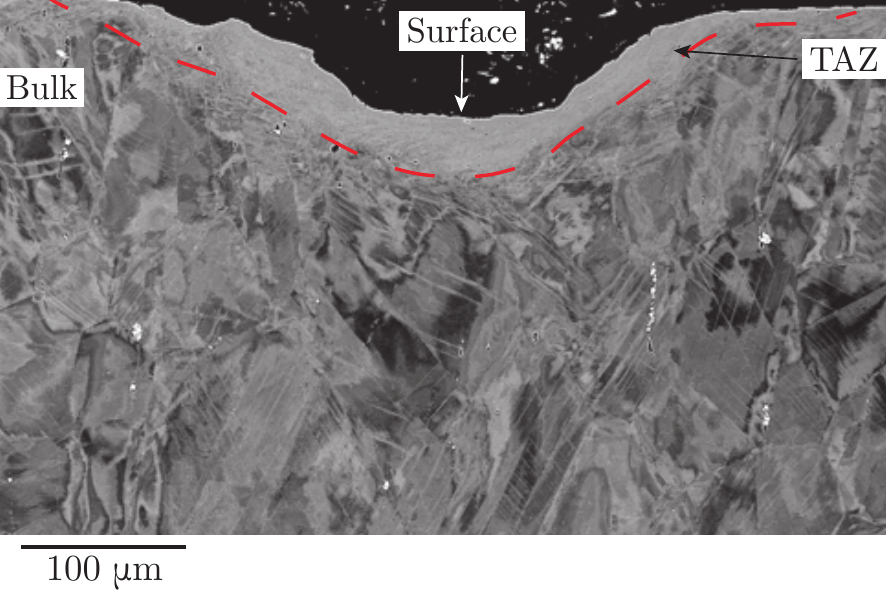}
			\caption{A radial cross-section of a galling trough in an unoxidised 316L stainless steel sample tested at a normal load of \SI{4.1}{\mega\pascal}.}
			\label{Rxsec}
		\end{figure}

		By sectioning galled samples radially, the subsurface deformation can be easily seen, Figure \ref{Rxsec}. The most apparent change in microstructure from the as-received material is the creation of a layer of material immediately beneath the gall scar. This layer is often referred to as the `white layer', `tribologically transformed zone' or `tribologically transformed structure' \cite{TribologyBook}. Since a transformation in phase or atomic structure may not necessarily occur, in this paper, this region is termed the `tribologically affected zone' (TAZ), analogous with the heat affected zone (HAZ) in welding, and will be discussed at length later in this paper. When viewed in a radial cross-section using an SEM, the structure of the TAZ is difficult to interpret from BSE images. The size of the trough depth and the TAZ depth beneath, can however be observed and noted as being \SI{67}{\micro\meter} and \SI{39}{\micro\meter} respectively, Figure \ref{Rxsec}, demonstrating that a significant damage layer is observed in galled samples.

			\begin{figure*}[h] %Where * allows figure environment to go over multiple columns
				\centering
				\includegraphics[width=18cm]{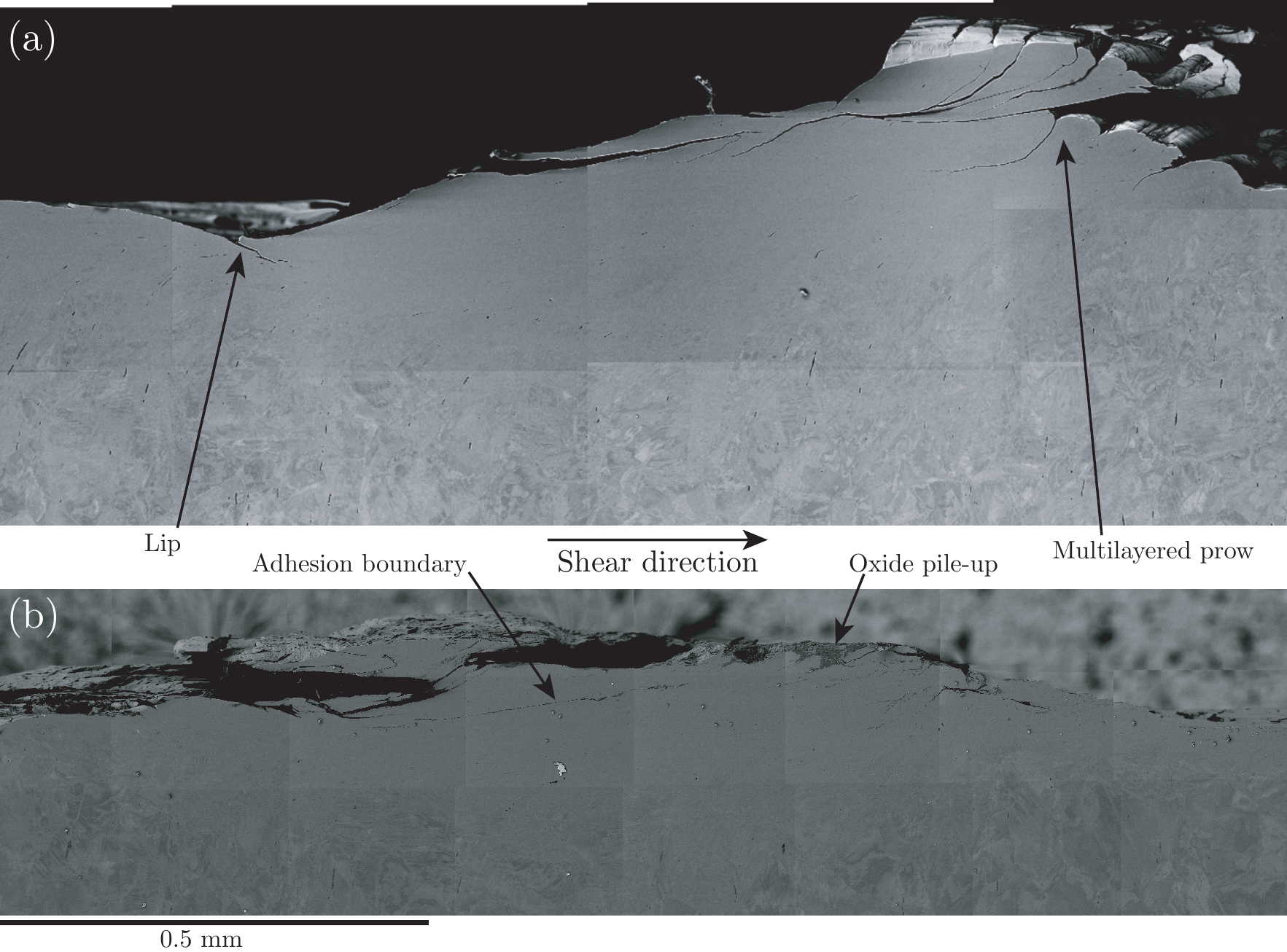}
				\caption{Circumferential cross-sections through gall scars in 316L stainless steel. (a) An unoxidised sample tested at \SI{3.8}{\mega\pascal}, where lips formed through shear failure and multiple layers within the prow are clearly visible. (b) An oxidised sample tested at \SI{93.6}{\mega\pascal}, where an adhesion boundary is visible due to the contrast of the oxide with the underlying stainless steel, and oxide pile-up on the sample surface is seen. Both samples show a region immediately beneath the sample surface where the microstructure is no longer visible; the tribologically affected zone (TAZ). The scale of both images is common.}
				\label{Circxsec}
			\end{figure*}
			
			\begin{figure}[t]
				\centering
				\includegraphics[width=9cm]{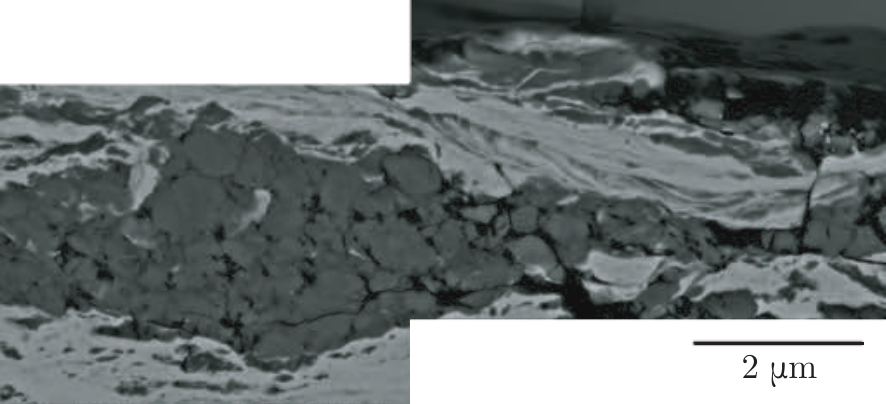}
				\caption{Fine scale mechanical mixing of the surface oxide layers (dark) with the underlying austenitic substrate (light).}
				\label{Steak}
			\end{figure}
			
			\begin{figure*}[t]
				\centering
				\includegraphics[width=18cm]{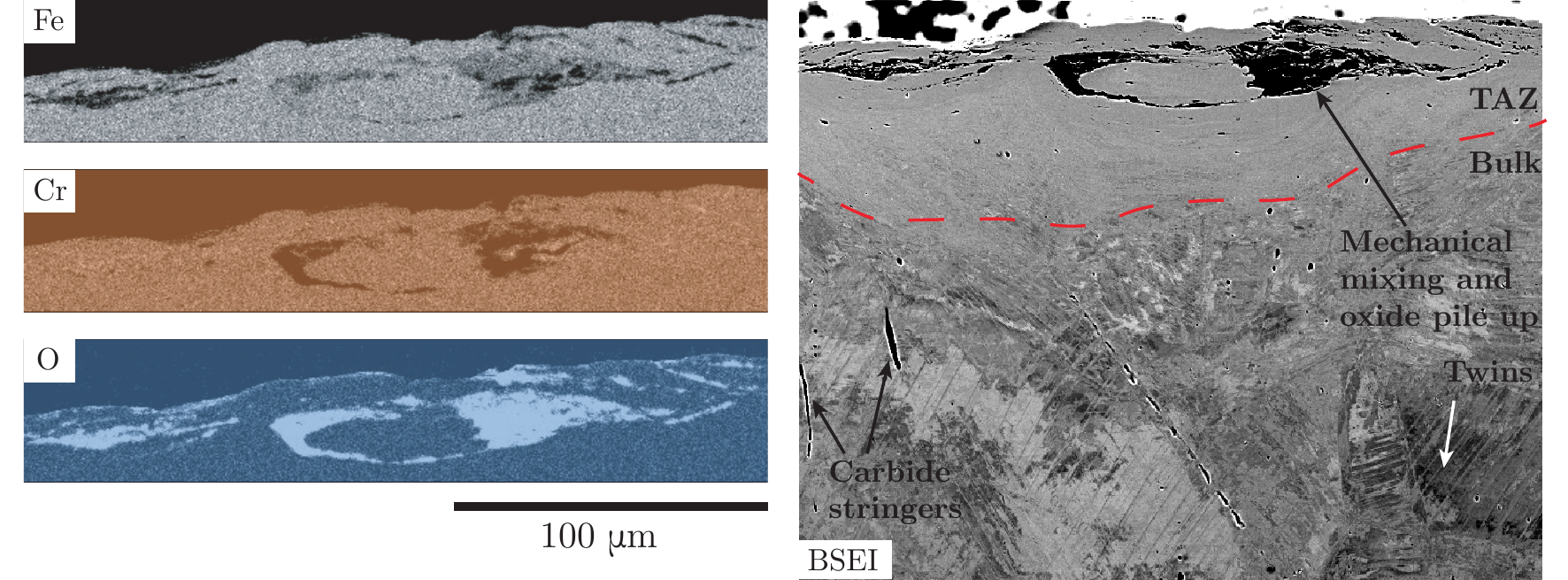}
				\caption{SEM-EDX imaging of a mechanically mixed region immediately beneath the sample surface. Both oxide layers are shown to be present by the chemical variation within the differing regions. The red box denotes the region examined using EDX.}
				\label{SEMEDX}
			\end{figure*}
			
		When viewed in the circumferential plane the mechanism is easier to rationalise, Figure \ref{Circxsec}. Although more obvious in the unoxidised sample, lips are seen to have formed within the galling trough, Figure \ref{Circxsec}. These lips are observed to be free of oxide. Similarly, in both the unoxidised and oxidised states a large multilayered prow is seen. In the oxidised state, these prows are not made of discrete layers, separated by a partially worn oxide surface, instead, having regions of mechanical mixing between the oxide layers and stainless steel substrate. This was particularly seen in radial cross-sections, where fine-scale mechanical mixing within peaks was observed, Figures \ref{Steak} \& \ref{SEMEDX}. The prows are therefore unlikely to have formed through the accumulation of plucked material, instead, forming through shear. This is consistent with the formation of the lips, which are known to be formed through shear failure \cite{VingsboMechanisms}. It is likely that as the prow grew it folded, and gave the appearance of a layered prow. This is particularly evident in the unoxidised sample at the front of the prow, where the surface is observed to have buckled, Figure \ref{Circxsec}(a). This can also be evidenced in the oxidised sample through the adhesion boundaries which are seen to contain a relatively large proportion of oxide, which appear not to be in intimate contact, Figure \ref{Circxsec}(b).
		
		In order for the initial adhesion in the oxidised sample to take place, a metal-metal contact must first be achieved. This is due to oxide-oxide and oxide-metal adhesion bonds being very weak \cite{AndersonSummary, 10YearStudy}. This therefore means that abrasion and removal of the oxide layers on both mating surfaces must occur before adhesion and galling may take place. The removal of the oxide layers therefore occurs through abrasion, with the abraded oxide being deposited in valleys within the surface, or within sample folds, Figure \ref{Circxsec}(b).
		
		Figure \ref{SEMEDX} shows a radial cross-section of a galling peak on an oxidised sample. The BSE image shows that within the bulk material there is extensive twinning, however, these were present in the as-received material. Twins were identified and distinguished from deformation-induced $\epsilon$-martensite by EBSD analysis. Immediately beneath the sample surface, oxide pile up as well as mechanical mixing of the surface, including oxides, can be seen, Figure \ref{SEMEDX}. By performing EDX analysis it can be observed that both oxides have been both mechanically mixed and piled up, Figure \ref{SEMEDX}. Between the region of mechanical mixing and the bulk, the TAZ is observed, however, it is difficult to interpret the microstructure of the TAZ using BSE in radial cross-sections. However, when observing the TAZ in a circumferential cross section, flow lines can be seen. In addition, although when viewed in radial cross-sections, carbide stringers appear unchanged from the as-received material, in circumferential cross-sections the carbide stringers are observed to follow flow lines, Figure \ref{Circxsec}(a). Since the carbide stringers are arranged perpendicular to the sample surface before-testing, by observing their post-test positions, it can be seen that extensive sub-surface shear has taken place, most notably by the shear lip, Figure \ref{Circxsec}(a).

			\begin{figure}[t]
				\centering
				\includegraphics[width=9cm]{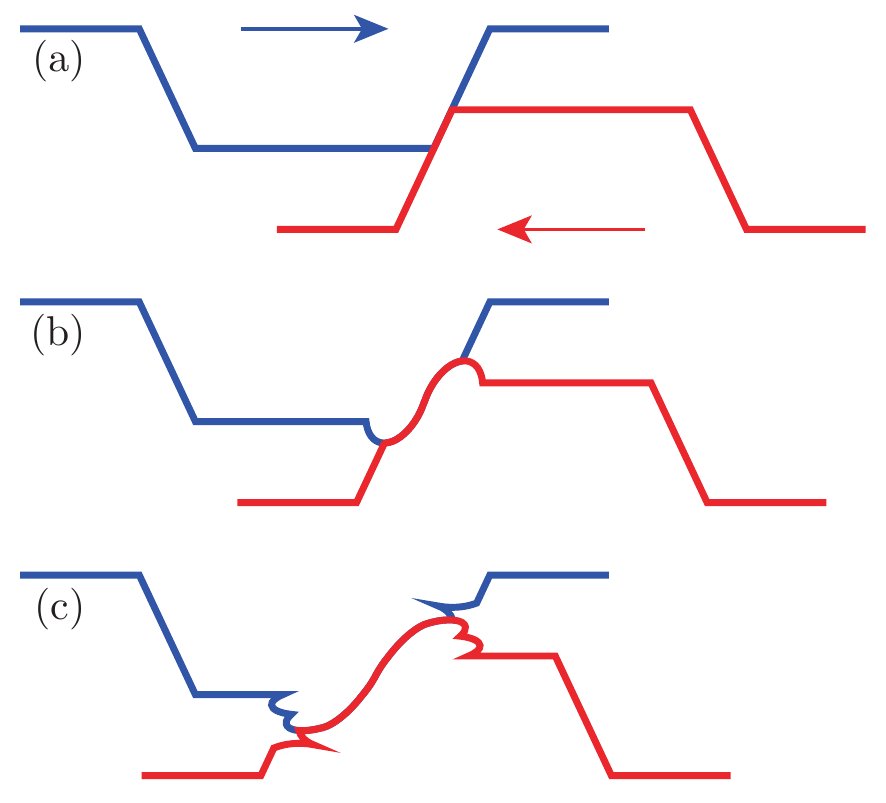}
				\caption{The wedge formation \& growth galling mechanism seen in 316L stainless steel. (a) Two asperities come into contact and form an adhesive junction; (b) shearing of this junction results in wedge formation (this can also occur through the shearing of two flat surfaces that have adhered); (c) the wedge grows to such an extent that excess material ahead of the prows folds over, whilst shear failure occurs behind the prow, resulting in the formation of lips. }
				\label{Mechanism}
			\end{figure}
			
		A number of galling mechanisms have been reported in the literature, however, only one of these was observed in the galling of 316L stainless steel; wedge formation and flow, Figure \ref{Mechanism} \cite{Antler1964,Cocks1964}. The wedge formation and flow mechanism is consistent with the observations of galled 316L stainless steel in the unoxidised condition by Peterson \textit{et al.} \cite{PetersonSummary}.

			Initially, adhesion of opposing surfaces takes place, either as two flat sections, in which case shear subsequently takes place to form a wedge \cite{Antler1964,Cocks1964}, or two asperities come into contact, essentially being pre-made wedges, Figure \ref{Mechanism}(a). Shear then continues to take place, causing material flow and wedge growth, Figure \ref{Mechanism}(b). As the prow continues to grow and is pushed from behind, the leading face of the prow will eventually fold over \cite{InSituStickSlip}, causing an additional interface within the prow, which, due to the compressive stresses it is under, will likely form an adhesion junction. Simultaneously, the trailing face of the prow continues to move, shearing the sub-surface material, and resulting in the formation of `lips' from shear failure, within the galling trough, \ref{Mechanism}(d) \cite{VingsboMechanisms}. This then continues such that multiple folds are formed as the wedge grows, whilst additional `lips' are seen, Figure \ref{Mechanism}.
			
		%HELPFUL INFORMATION:
		
		%%FULL MECH SEEN IN GALLING BOOK p77 (including peak formation and lip formation in gall scar)
		
		%%Adhesion initially occurs between 'flat' surfaces as per \cite{Cocks1962,Macdonald1971} [...nominally flat surfaces. These 'flat surfaces...]
			%THEN
		%%Kayaba (\cite{SuccessiveWear} uses 304 (+ others)
		%\cite{VingsboMechanisms} Fig 3c shows shear fracture at adhesive junctions - show this INCLUDE: micrograph showing this (surface) + label in UnOx Circ
			%%|->we therefore see the mechanisms which results in the layers seen in the peaks, whilst forming multiple layers through mechanism shown by Kayaba
			%|-> p627 (top) explicitly states that corrosion which result in oxides that can act as lubricants i.e. corrosion causes lubrication
			
		%%\cite{SteelersGalling} 316 known to gall at low stresses + detailed look at 316 in \cite{PetersonSummary} [conclusion 3 = "making & breaking of welds"]

	\subsection{Tribologically Affected Zone}
		Something which has been studied very little in the literature is the sub-surface structure which results from galling. It is known that immediately beneath the galling scar, the hardness of the material is increased \cite{KimKim}, particularly in the tribologically affected zone (TAZ). 
				
		\begin{figure}[h]
				\centering
				\includegraphics[width=9cm]{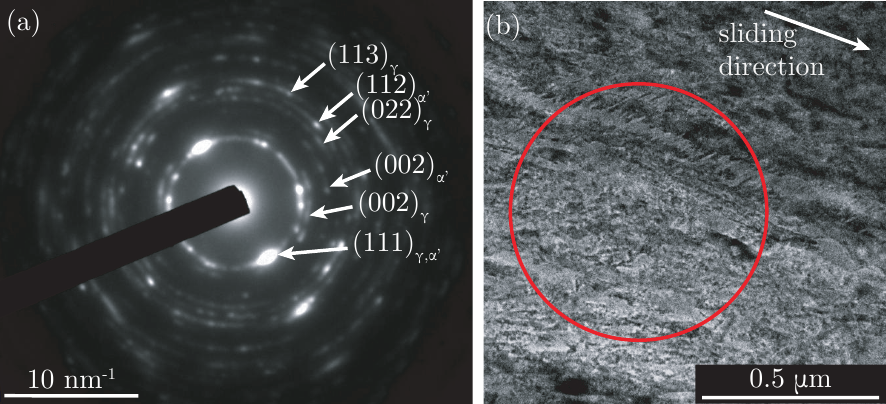}
				\caption{(a) A TEM diffraction pattern of the region shown in (b), which is a portion of the tribologically affected zone immediately beneath a galling trough. An \textit{in-situ} FIB lift-out was taken from an unoxidised sample which had been galled at \SI{4.1}{\mega\pascal} and radially cut.}
				\label{TAZ TEM}
			\end{figure}

		TEM imaging was used to discern the microstructure of the TAZ by removing a section of the TAZ from a radial cross-section  with a FIB. The TAZ was observed to be a nanocrystalline region where grains are elongated in the shear direction, Figure \ref{TAZ TEM}. Since the sample was nanocrystalline, diffraction rings were formed, and the distance of these from the straight-through beam were measured and their crystal planes indexed. The indexing of these rings showed the presence of bcc-ferrite. This is surmised to be low-carbon (and hence low tetragonality) deformation-induced bct martensite formed without diffusion and compositional change. This conclusion was verified using X-ray diffraction (XRD) and STEM-EDX.
			
			\begin{figure}[h]
				\centering
				\includegraphics[width=9cm]{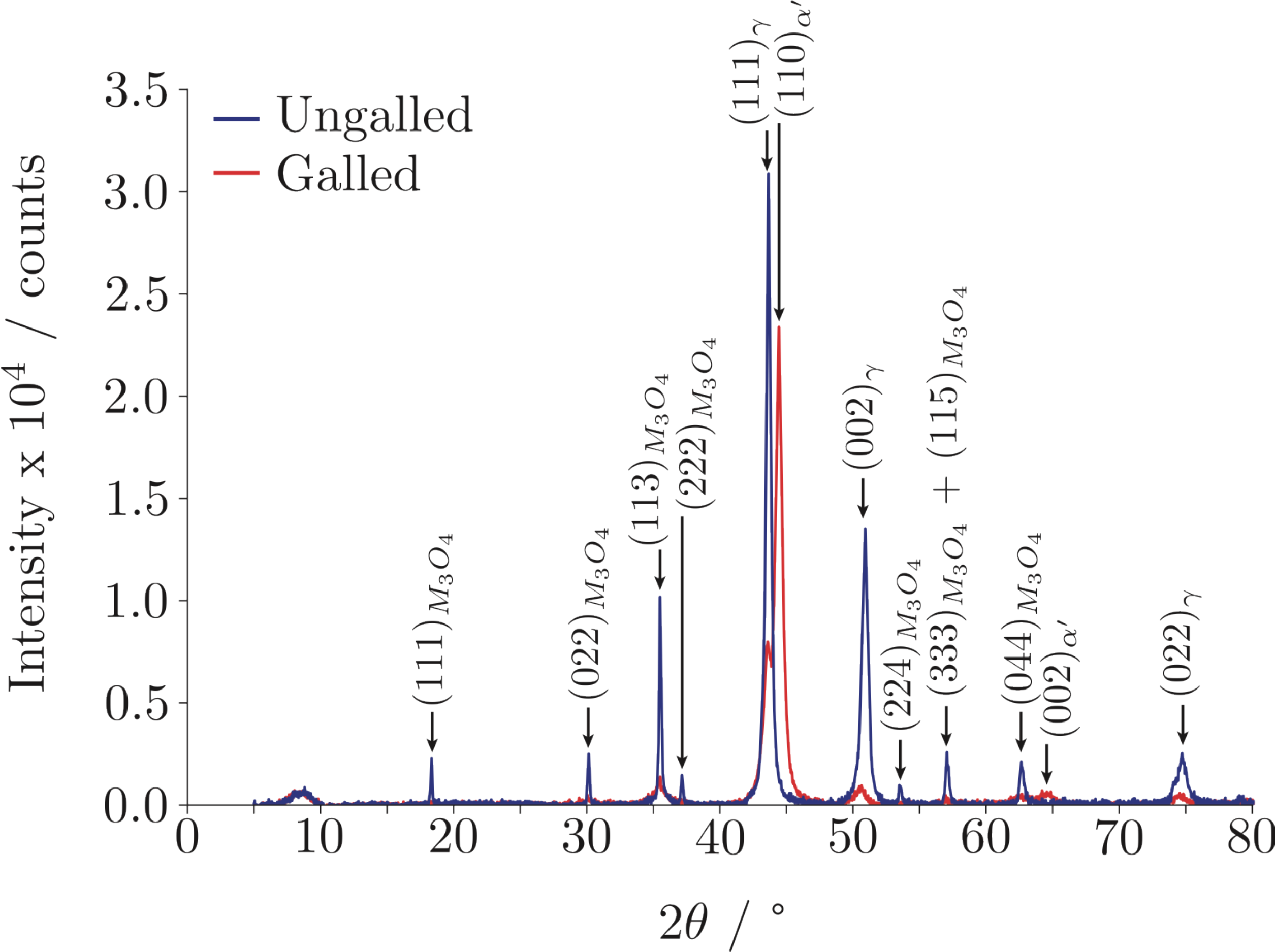}
				\caption{X-ray diffraction patterns of samples which had been oxidised in simulated autoclave conditions and where one sample was then galled at \SI{96.5}{\mega\pascal}.}
				\label{Ox XRD}
			\end{figure}
		
		Figure \ref{Ox XRD} shows the XRD patterns obtained for oxidised 316L both pre- and post-galling and additional peaks are seen, which are in positions corresponding to a bcc-phase. STEM-EDX found that the concentration of Mn (an austenite stabiliser), was constant across the sample, verifying the fact that a low-tetragonality bct-martensite phase has formed as a result of the extensive shear experienced immediately below the galling surface.
		
		In the literature, a number of authors consider such transformation induced plasticity (TRIP) behaviour to be beneficial in galling resistance, however, this is yet to be confirmed. \cite{SmithThesis,BuckleyHex}.

\section{Conclusion}
	In this work, 316L stainless steel was oxidised in simulated PWR conditions, forming a multilayer oxide. Both oxide layers were found, using XRD, to be magnetite, M$_{3}$O$_{4}$. Using STEM and STEM-EDX, the outer oxide was found to have a composition of Fe$_{3}$O$_{4}$, and was seen to have formed discrete single crystals across the sample surface. The inner oxide was found to be a continuous nanocrystalline layer which was Cr-rich, being of approximate composition Cr$_{1.3}$Fe$_{1.2}$Ni$_{0.5}$O$_{4}$.
	
	When galled in both the unoxidised and oxidised. conditions, 316L stainless steel is seen to gall via the wedge formation \& flow mechanism. In order for this mechanism to occur, however, sufficient oxide must be removed to enable an adhesion bond stronger than a cohesion bond local to the adhesion surface. This therefore means that abrasion or mechanical mixing must first occur before gall can take place when 316L is in the oxidised state, since oxide-oxide \& metal-oxide adhesion bonds are significantly weaker than metal-metal adhesion bonds. For this reason, the galling performance of 316L stainless steel can be improved by around 30x (\SI{780}{\micro\meter} \textit{vs.} \SI{26}{\micro\meter} under a normal load of \SI{4.2}{\mega\pascal}) by forming an oxide scale.
	
	A multilayered peak and a trough, with `lips', indicative of shear fracture are formed during the galling and have been recorded as giving a sample R$_{t}$ of up to \SI{0.8}{\milli\meter}. It has been observed that in both the oxidised and unoxidised states, an increase in normal load correspond to an increase in galling damage, as recorded using R$_{t}$ and galling area.
	
	Since shear failure occurs during the wedge formation \& flow mechanism, it is unsurprising that immediately beneath the sample surface, a region of extensive shear is observed, named the tribologically affected zone, TAZ. The TAZ has been found to be nanocrystalline, being a mixture of parent austenite and martensite, formed during the shear deformation.

	%Previously written section	
	%The adhesion of mated surfaces results in extensive sub-surface shear. This shear can result in prow and trough formation, the pile-up and removal of material, respectively. In regions that have been heavily shear and have not failed, extensive deformation is seen as shear flow and the formation of strain-induced martensite.
	
	%Both prow and trough formation are observed to have formed through successive stick-slip. This is evidenced by 'lips' seen in troughs and the layered structure seen in peaks.

\section*{Acknowledgements}
We gratefully acknowledge support from Rolls-Royce plc, from EPSRC (EP/N509486/1, EP/N025954/1 and EP/R000956/1) and from the Royal Society (D Dye Industry Fellowship).

\section*{References}	%* denotes lack of number
\bibliography{Paper1Refs}

\begin{thebibliography}{10}

\bibitem{ONRCobalt}
M.~Vannerem.
\newblock Chemistry of operating civil nuclear reactors.
\newblock ONR Guide NS-TAST-GD-088 Revision 2, Office for Nuclear Regulation,
  January 2019.

\bibitem{RTOcken}
H.~Ocken.
\newblock The galling wear resistance of new iron-base hardfacing alloys: a
  comparison with established cobalt- and nickel-base alloys.
\newblock {\em Surface and Coatings Technology}, 76-77:456--461, 1995.

\bibitem{NOREM02Friction}
D.H.E.~Persson, S.~Jacobson and S.~Hogmark.
\newblock Effect of temperature on friction and galling of laser processed
  norem 02 and stellite 21.
\newblock {\em Wear}, 255(1):498--503, 2003.

\bibitem{BowdenThesis}
D.~Bowden.
\newblock {\em Assessment of cobalt-free hardfacing stainless steel alloys for
  nuclear applications}.
\newblock PhD thesis, University of Manchester, 2016.

\bibitem{GallTough}
J.H.~Magee.
\newblock Two galling resistant stainless steels used for bridge hinge pins.
\newblock Technical report, Carpenter Technology Corporation, 2018.

\bibitem{SteelersGalling}
Committee of~Stainless Steel~Producers.
\newblock Review of the wear and galling characteristics of stainless steels.
\newblock Handbook.

\bibitem{HardfacingIndustrialPractices}
I.~Inglis, E.V.~Murphy and H.~Ocken.
\newblock Performance of wear-resistant iron base hardfacing alloys in valves
  operating under prototypical pressurized water reactor conditions.
\newblock {\em Surface and Coatings Technology}, 53:101--106, 1992.

\bibitem{TribologyBook}
I.~Hutchings and P.~Shipway.
\newblock {\em Tribology: Friction and Wear of Engineering Materials}.
\newblock Butterworth-Heinemann, second edition edition, 2017.

\bibitem{Cocks1958}
M.~Cocks.
\newblock Wear debris in the contact between sliding metals.
\newblock {\em Journal of Applied Physics}, 29:1609--1610, 1958.

\bibitem{Antler1962}
M.~Antler.
\newblock Wear, friction, and electrical noise phenomena in sever sliding
  systems.
\newblock {\em ASLE Transactions}, 5:297--307, 1962.

\bibitem{Antler1964}
M.~Antler.
\newblock Processes of metal transfer and wear.
\newblock {\em Wear}, 7(2):181--203, 1964.

\bibitem{Sasada1976}
T.~Sasada and S.~Norose.
\newblock The formation and growth of wear particles through mutual material
  transfer.
\newblock In {\em Proceedings of the JSLE-ASLE International Lubrication
  Conference}, p 82--91, 1976.

\bibitem{Cocks1962}
M.~Cocks.
\newblock Interaction of sliding metal surfaces.
\newblock {\em Journal of Applied Physics}, 33(7):2152--2161, 1962.

\bibitem{Cocks1964}
M.~Cocks.
\newblock Role of displaced metal in the sliding of flat metal surfaces.
\newblock {\em Journal of Applied Physics}, 35(6):1807--1814, June 1964.

\bibitem{Glaeser1981}
W.A.~Glaeser.
\newblock Wear experiments in the scanning electron microscope.
\newblock {\em Wear}, 73(2):371--386, 1981.

\bibitem{StickSlip}
T.~Kayaba, K.~Kato and Y.~Nagasawa.
\newblock Abrasive wear in stick-slip motion.
\newblock In {\em Wear of Materials: International Conference on Wear of
  Materials, 1981}, p 439--446, 1981.

\bibitem{SuccessiveWear}
T.~Kayaba and K.~Kato.
\newblock The analysis of adhesive wear mechanism by successive observation of
  the wear process in sem.
\newblock In {\em Wear of Materials}, p 45--56, April 1979.

\bibitem{SmithThesis}
R.T.~Smith.
\newblock {\em Development of a Nitrogen-Modified Stainless-Steel Hardfacing
  Alloy}.
\newblock PhD thesis, The Ohio State University, 2015.

\bibitem{KimKim}
J.-K.~Kim and S.-J.~Kim.
\newblock The temperature dependence of the wear resistance of iron-base NOREM 02 hardfacing alloy.
\newblock {\em Wear}, 237(2):217--222, February 2000.

\bibitem{TTSFrettingMechanisms}
E.~Sauger, L.~Ponsonnet, J.M.~Martin and L.~Vincent.
\newblock Study of the tribologically transformed structure created during
  fretting tests.
\newblock {\em Tribology International}, 33(11):743--750, 2000.

\bibitem{TTSFrettingSteel}
V.~Nurmi, J.~Hintikka, J.~Juoksukangas, M.~Honkanen, M.~Vippola, A.~Lehtovaara, A.~Mantyla, J.~Vaara and T.~Frondelius.
\newblock The formation and characterization of fretting-induced degradation
  layers using quenched and tempered steel.
\newblock {\em Tribology International}, 131:258--267, March 2019.

\bibitem{ASTMG98}
ASTM International.
\newblock ASTM G98-17 standard test method for galling resistance of materials.
\newblock Technical report, ASTM International, 2017.

\bibitem{ASTMG196}
ASTM International.
\newblock ASTM G196-08(2016) standard test method for galling resistance of
  material couples.
\newblock Technical report, ASTM International, 2016.

\bibitem{GallingMatrix}
K.G.~Budinski and S.T.~Budinski.
\newblock Interpretation of galling tests.
\newblock {\em Wear}, 332(C):1185--1192, 2015.

\bibitem{GallingBook}
L.K.~Ives, M.B.~Peterson and E.P.~Whitenton.
\newblock The mechanism, measurement, and influence of properties on the
  galling of metals.
\newblock Technical Report NISTIR 89-4064, National Institute of Standards and
  Technology (NIST), December 1989.

\bibitem{CoLWR}
H.~Ocken.
\newblock Reducing the cobalt inventory in light water reactors.
\newblock {\em Nuclear Technology}, 68:18--28, January 1985.

\bibitem{CorrosionPWRWaterChemistry}
T.~Terachi, T.~Yamada, T.~Miyamoto, K.~Arioka and K.~Fukuya.
\newblock Corrosion behavior of stainless steels in simulated pwr primary
  water---effect of chromium content in alloys and dissolved hydrogen.
\newblock {\em Journal of Nuclear Science and Technology}, 45(10):975--984,
  2008.

\bibitem{Kim316}
Y.-J.~Kim.
\newblock Characterization of the oxide film formed on type 316 stainless steel
  in 288$\,^{\circ}$C water in cyclic normal and hydrogen water chemistries.
\newblock {\em Corrosion}, 51(11):849--860, 1995.

\bibitem{BuckleyHex}
D.H.~Buckley and R.L.~Johnson.
\newblock Friction and wear of hexagonal metals and alloys as related to
  crystal structure and lattice parameters in vacuum.
\newblock {\em ASLE Transactions}, 9(2):121--135, 1966.

\bibitem{BuckleySummary}
D.H.~Buckley.
\newblock Surface films and metallurgy related to lubrication and wear.
\newblock {\em Progress in Surface Science}, 12(1):1--154, 1982.

\bibitem{VingsboMechanisms}
O.~Vingsbo.
\newblock Wear and wear mechanisms.
\newblock In {\em Wear of Materials}, p 620--635. ASME, 1979.

\bibitem{AndersonSummary}
O.L.~Anderson.
\newblock Adhesion of solids: Principles and applications.
\newblock {\em Bell Laboratories Record}, 35(11):441--445, 1957.

\bibitem{10YearStudy}
S.B.~Ainbinder and A.S.~Fran{\v c}s.
\newblock On the mechanism of the formation and destruction of adhesion
  junctions between bodies in frictional contact.
\newblock {\em Wear}, 9(3):209--227, 1966.

\bibitem{PetersonSummary}
 M.~Peterson, K.J.~Bhansali, E.P.~Whitenton and L.K.~Ives.
\newblock Galling wear of materials.
\newblock In {\em American Society of Mechanical Engineers Wear of Materials:
  International Conference: Papers}, p 293--301, 1985.

\bibitem{InSituStickSlip}
H.~Yeung, K.~Viswanathan, A.~Mahato and S.~Chandrasekar.
\newblock Surface phenomena revealed by in situ imaging: studies from adhesion,
  wear and cutting.
\newblock {\em Surface Topography: Metrology and Properties}, 5(1):014002,
  2017.

\end{thebibliography}
	
\end{document}